\documentclass[12pt]{article}
\usepackage{graphicx}
\usepackage{amssymb,amsmath}
\usepackage{times}

\begin{document}

\title{Multicolour Optical Surface Brightness Profiles Decomposition of
the Seyfert \\ Galaxies III~Zw~2, Mrk~506 and Mrk~509\thanks{Based on
observations obtained at the Rozhen National Astronomical Observatory
of Bulgaria operated by the Institute of Astronomy, Bulgarian Academy
of Sciences.}}
\author{L. Slavcheva-Mihova, B. Mihov, G. Petrov and V. Kopchev \\ \\
{\em Institute of Astronomy, Bulgarian Academy of Sciences,} \\
{\em 72 Tsarigradsko Chausse Blvd., 1784 Sofia, Bulgaria} \\
{\tt lslav,bmihov,petrov@astro.bas.bg}}
\date{May 2006}
\maketitle

\begin{abstract}
We present the results of the $UBVR_{\rm \scriptscriptstyle C}I_{\rm \scriptscriptstyle C}$
surface brightness profiles decomposition of the Seyfert galaxies III~Zw~2, Mrk~506 and Mrk~509.
The profiles were modelled as a sum of a Gaussian, a S\'ersic law and an exponent. A Ferrers bar
and a Gaussian ring were added to the model profiles of III~Zw~2 and Mrk~506, respectively. The
parameters and the total magnitudes of the structural components were derived.
\end{abstract}

\section{Introduction}

We present here the first results of an outgoing study aimed to do a detailed decomposition
of Seyfert galaxies $UBVR_{\rm \scriptscriptstyle C}I_{\rm \scriptscriptstyle C}$ surface
brightness profiles (hereafter SBPs). Our advantages over Seyfert galaxies SBP decomposition
found in the literature are the following: (1) we model explicitly the active nucleus in
opposition to some authors who avoid nucleus modelling (e.g. \cite{chatzichristou01}),
(2) we use S\'ersic rather than de~Vaucouleurs law for bulge, (3) we use a truncated exponential
law defined in \cite{kormendy77} along with a pure exponent in disk modelling, and (4) we
model bar/oval/lens/ring components that have been generally skipped by Seyfert galaxies SBPs
decomposers (e.g. \cite{boris02}).

All galaxies to be decomposed were observed at Rozhen NAO of Bulgaria with the 2-m telescope
and Photometrics AT200 CCD camera ($0.309\,\rm arcsec\,\rm px^{-1}$) through a standard
Johnson-Cousins $UBVR_{\rm \scriptscriptstyle C}I_{\rm \scriptscriptstyle C}$ set of filters.
The SBPs of the galaxies were extracted fitting ellipses to the galaxian isphotes by means of
${\rm \scriptstyle FIT/ELL3}$ command of the ${\rm \scriptstyle SURFPHOT}$ context of
${\rm \scriptstyle ESO-MIDAS}$ package (see \cite{slavcheva05} for details).

Our first decomposition results concern the Seyfert galaxies III~Zw~2 (Sy1.0),
Mrk~506 (Sy1.5) and Mrk~509 (Sy1.2); the Seyfert types are taken from NED.

\section{Models and Methods}

Our basic model SBP is a sum of (1) a Gaussian with a fixed FWHM to represent the nucleus
(the only free parameter is the central surface brightness), (2) a S\'ersic law \cite{sersic68}
with free parameters $\mu_{\rm eff}$ -- the effective surface brightness, $r_{\rm eff}$ -- the
effective radius, and $n$ -- the power-law index, to represent the bulge and (3) an exponent \cite{freeman70}
with free parameters $\mu_{\rm cen}$ -- the central surface brightness, and $r_{\rm scl}$ -- the scale
length, to represent the disk. Bulge and disk model SBPs were convolved with a Gaussian PSF
to simulate the seeing effects on the profiles according to \cite{bendinelli82,bailey83}.
Note that if the frame PSF is not circular then the FWHM of the convolution Gaussian is
set to the PSF's mean FWHM along the minor axis and the FWHM of the nuclear Gaussian is set
to the PSF's mean FWHM along the major axis (the mean FWHM of the stellar images for each
frame was determined fitting a 2D Gaussian to a number of field stars employing
${\rm \scriptstyle CENTER/IQE}$ command within ${\rm \scriptstyle ESO-MIDAS}$).

We added (1) a Ferrers model profile \cite{laurikainen05} with free parameters $\mu_{\rm cen}$ -- the
central surface brightness, $r_{\rm end}$ -- the profile length, and $m$ -- the power-law index,
to account for the bar\footnote{The bar manifests in about 0.25 rise of the
ellipticity and an almost constant position angle.} in III~Zw~2, and (2) a displaced
Gaussian model profile \cite{prieto01} with free parameters $\mu_{\rm cen}$ -- the central surface
brightness, $r_{\rm cen}$ -- the position (or displacement) of the Gaussian centre, and
$\cal FW$ -- the FWHM, to account for the ring in Mrk~506 (the galaxy has {\em SAB(r)a}
morphology according to NED).

The figure-of-merit function that is minimized is equal to the unweighted sum of the squared differences
between the observed and the model SBPs per degree of freedom, $\nu$:
\begin{equation}
\label{delta}
\Delta^2_{\nu}(\textbf{\textit{p}})=\nu^{-1}\,\sum_{i=1}^N {\Big [}\mu^{\,\rm \scriptscriptstyle OBS}_i - \mu^{\,\rm \scriptscriptstyle MOD}_i(\textbf{\textit{p}}){\Big ]}^2,
\end{equation}
where $\textbf{\textit{p}}$ is the $P$-element vector of the free model parameters to be fitted.
The initial value of the degrees of freedom is defined as $\nu_0=N-P$, where $N$ and $P$ are
the number of the profile data points and the number of the free parameters, $\textbf{\textit{p}}$,
to be fitted, respectively. Note that corrections to $\nu_0$ could be done for the
presence of (1) zero-weighted profile data points, $N_{\rm zero}$, and/or (2) fixed parameters,
$P_{\rm fix}$, so the actual value of $\nu$ becomes $\nu=\nu_0-N_{\rm zero}+P_{\rm fix}$,
and the corrected value of $\nu$ enters Eq.~\ref{delta}. The minimization of
$\Delta^2_{\nu}(\textbf{\textit{p}})$ was performed employing Levenberg-Marquardt algorithm \cite{press92}.
The initial guess parameters were estimated by eye overplotting the observed and the model
profiles and changing the parameters manually to get a good correspondence between them. After that,
a $P$-dimensional {\em valley} $[\textbf{\textit{p}}-\delta \textbf{\textit{p}}^-,\,\textbf{\textit{p}}+\delta \textbf{\textit{p}}^+]$
around the initial guess parameters was constructed; here $\delta \textbf{\textit{p}}^{+/-}$
are the allowed parameters' deviations from their initial guess values in both directions
(these deviations could be different for each parameter). Next, a number of decomposition cycles
were run: in each cycle the actual initial guess parameters were picked up randomly from the
uniformly distributed parameters in the {\em valley} defined above. This procedure helped us in
isolating the global minimum among $\Delta^2_{\nu}(\textbf{\textit{p}})$ values -- the larger
the number of the random cycles is, the bigger the probability to find the global minimum becomes. The
number of random cycles could vary depending on the complexity of the profile to be
decomposed -- more complicated profiles could require up to several hundreds of random cycles.
After the minimization cycles were completed a histogram of $\Delta^2_{\nu}(\textbf{\textit{p}})$
values was built and the minimum corresponding to the most frequently occurring
$\Delta^2_{\nu}(\textbf{\textit{p}})$ was selected. If the parameters corresponding to this minimum
behaved themselves well and if this minimum had the lowest value of $\Delta^2_{\nu}(\textbf{\textit{p}})$
then it was assumed to be the global one with $\Delta^2_{\nu}(\textbf{\textit{p}}_{\rm min})=\Delta^2_{\nu,\,\rm min}$;
the correspondence between the parameter values obtained after decomposition of different
observing runs profiles could be used as a further check of the minimum found (see Table~\ref{pars}).
If some of the parameters corresponding to the most frequently occurring minimum had unacceptable
values (e.g. very small or very large), then this minimum was rejected and a new global minimum
was searched for among the remaining random cycles results. If there were minima with
$\Delta^2_{\nu}(\textbf{\textit{p}})$ values lower than $\Delta^2_{\nu}(\textbf{\textit{p}})$
of the most frequently occurring minimum then these minima were checked one by one; in all
cases we found that the parameters corresponding to these minima had unacceptable values,
i.e. these minima were local ones.

\section{Results}

At the end of the selected objects decomposition the best-fit parameters were obtained (listed
in Table~\ref{pars}) and, based on them, the total magnitudes of the structural components
were computed (listed in Table~\ref{mags}). We list only the total magnitudes for the nuclear
Gaussian because its central surface brightness is strongly dependent on the seeing and
does not allow straightforward comparison of the results obtained at nights with different
seeing conditions. The structural parameter(s) and/or the total magnitude(s) that could not
be derived from the decomposition are marked in the tables. The errors of the parameters are
$1\sigma$ uncertainties as resulting from the fitting algorithm and they should be considered
as approximate ones. More reliable estimate of the parameter errors could be obtained
through Monte Carlo simulations or a bootstrap analysis. The mean FWHM along the minor
axis of the stellar images, ${\overline {\cal FW}}{\rm \scriptscriptstyle PSF}$,
that was used in the model SBPs convolution and the values of
$\sigma_{\rm fit}=(\Delta^2_{\nu,\,\rm min})^{0.5}$ are listed in Table~\ref{mags} as
well. The civil date and the name of the decomposed object are listed in the tables using
the following code: 1a -- September 9/10, 1997 III~Zw~2; 1b -- June 1/2, 1997 Mrk~506;
2b -- July 18/19, 1998 Mrk~506; 1c -- July 10/11, 1997 Mrk~509; 2c -- September 8/9, 1997 Mrk~509;
3c -- July 20/21, 1998 Mrk~509. Corresponding Johnson-Cousins filter is shown along with the date
and the object code. We list the parameters of the bar in III~Zw~2 and of the ring in Mrk~506 in
Table~\ref{bar} and Table~\ref{ring}, respectively. Note that the magnitudes and the surface
brightnesses listed in the tables have not been corrected for the Galactic absorption and
cosmological dimming; $k$- and evolution corrections have not been applied as well.

\begin{table}
\begin{center}
\caption{\label{pars} Bulge $(^{\rm \scriptscriptstyle B})$ and disk $(^{\rm \scriptscriptstyle D})$ structural
parameters for III~Zw~2, Mrk~506 and Mrk~509. The errors of the parameters are shown as a superscript index to
the corresponding values.}\smallskip
\begin{tabular}{lrrrrr}
\hline \noalign {\smallskip}
$\rm Code$ {\rule[-1.7ex]{0pt}{0pt}} & $\mu^{\,\rm \scriptscriptstyle B}_{\,\rm eff}$ & $r^{\,\rm \scriptscriptstyle B}_{\,\rm eff}$ & $n^{\,\rm \scriptscriptstyle B}$ & $\mu^{\,\rm \scriptscriptstyle D}_{\,\rm cen}$ & $r^{\,\rm \scriptscriptstyle D}_{\,\rm scl}$ \\
{\rule[-1.7ex]{0pt}{0pt}} & $[\,\rm mag\,/\,\Box^{\,\prime \prime}\,]$ & $[\,^{\prime \prime}\,]$ & & $[\,\rm mag\,/\,\Box^{\,\prime \prime}\,]$ & $[\,^{\prime \prime}\,]$ \\
\hline \noalign {\smallskip}

$1{\rm a}B$                                                      & $18.762^{\,0.193}$ & $0.849^{\,0.058}$ & $0.976^{\,0.099}$ & $23.380^{\,0.028}$ & $5.138^{\,0.044}$ \\
$1{\rm a}V$                                                      & $18.182^{\,0.143}$ & $0.847^{\,0.047}$ & $1.117^{\,0.104}$ & $22.001^{\,0.017}$ & $5.732^{\,0.033}$ \\
$1{\rm a}R_{\rm \scriptscriptstyle C}$                           & $17.346^{\,0.315}$ & $0.784^{\,0.092}$ & $0.915^{\,0.186}$ & $21.146^{\,0.051}$ & $5.318^{\,0.085}$ \\
$1{\rm a}I_{\rm \scriptscriptstyle C}$ {\rule[-1.0ex]{0pt}{0pt}} & $17.939^{\,0.340}$ & $1.114^{\,0.138}$ & $0.615^{\,0.177}$ & $20.312^{\,0.023}$ & $5.766^{\,0.042}$ \\

\hline \noalign {\smallskip}

$2{\rm b}U$                                                      & $18.181^{\,0.210}$ & $1.003^{\,0.075}$ & $1.079^{\,0.140}$ & $21.404^{\,0.053}$ & $6.473^{\,0.124}$ \\
$1{\rm b}B$                                                      & $               -$ & $              -$ & $              -$ & $21.011^{\,0.010}$ & $6.222^{\,0.021}$ \\
$2{\rm b}B$                                                      & $19.825^{\,0.242}$ & $1.342^{\,0.153}$ & $0.316^{\,0.171}$ & $21.110^{\,0.023}$ & $6.185^{\,0.045}$ \\
$1{\rm b}V$                                                      & $               -$ & $              -$ & $              -$ & $20.227^{\,0.007}$ & $6.346^{\,0.013}$ \\
$2{\rm b}V$                                                      & $19.168^{\,0.107}$ & $1.292^{\,0.054}$ & $0.592^{\,0.077}$ & $20.172^{\,0.016}$ & $6.198^{\,0.031}$ \\
$1{\rm b}R_{\rm \scriptscriptstyle C}$                           & $19.563^{\,2.958}$ & $1.287^{\,1.324}$ & $1.192^{\,1.694}$ & $19.579^{\,0.013}$ & $6.051^{\,0.022}$ \\
$2{\rm b}R_{\rm \scriptscriptstyle C}$                           & $19.040^{\,0.043}$ & $1.473^{\,0.029}$ & $0.460^{\,0.036}$ & $19.628^{\,0.010}$ & $6.285^{\,0.018}$ \\
$1{\rm b}I_{\rm \scriptscriptstyle C}$                           & $19.619^{\,0.050}$ & $2.057^{\,0.054}$ & $0.647^{\,0.083}$ & $19.161^{\,0.009}$ & $7.086^{\,0.023}$ \\
$2{\rm b}I_{\rm \scriptscriptstyle C}$ {\rule[-1.0ex]{0pt}{0pt}} & $18.149^{\,0.034}$ & $1.298^{\,0.019}$ & $0.658^{\,0.036}$ & $18.939^{\,0.014}$ & $6.271^{\,0.030}$ \\

\hline \noalign {\smallskip}

$3{\rm c}U$                                                      & $16.915^{\,0.201}$ & $1.545^{\,0.062}$ & $1.203^{\,0.042}$ & $21.343^{\,0.058}$ & $5.995^{\,0.119}$ \\
$1{\rm c}B$                                                      & $18.156^{\,0.086}$ & $1.429^{\,0.045}$ & $1.359^{\,0.065}$ & $21.933^{\,0.045}$ & $5.763^{\,0.092}$ \\
$2{\rm c}B$                                                      & $17.184^{\,0.091}$ & $0.981^{\,0.032}$ & $1.800^{\,0.063}$ & $21.848^{\,0.015}$ & $6.425^{\,0.033}$ \\
$3{\rm c}B$                                                      & $17.417^{\,0.212}$ & $1.450^{\,0.074}$ & $1.108^{\,0.044}$ & $21.411^{\,0.022}$ & $5.546^{\,0.042}$ \\
$1{\rm c}V$                                                      & $18.443^{\,0.066}$ & $1.880^{\,0.047}$ & $0.850^{\,0.044}$ & $20.514^{\,0.018}$ & $5.046^{\,0.025}$ \\
$2{\rm c}V$                                                      & $18.210^{\,0.082}$ & $1.585^{\,0.049}$ & $1.330^{\,0.075}$ & $20.910^{\,0.027}$ & $6.654^{\,0.070}$ \\
$3{\rm c}V$                                                      & $17.414^{\,0.144}$ & $1.400^{\,0.070}$ & $1.508^{\,0.096}$ & $20.877^{\,0.028}$ & $6.253^{\,0.060}$ \\
$1{\rm c}R_{\rm \scriptscriptstyle C}$                           & $17.765^{\,0.034}$ & $1.717^{\,0.023}$ & $1.053^{\,0.028}$ & $20.051^{\,0.011}$ & $5.306^{\,0.017}$ \\
$2{\rm c}R_{\rm \scriptscriptstyle C}$                           & $17.744^{\,0.049}$ & $1.633^{\,0.032}$ & $1.405^{\,0.049}$ & $20.550^{\,0.016}$ & $6.841^{\,0.035}$ \\
$3{\rm c}R_{\rm \scriptscriptstyle C}$                           & $17.886^{\,0.072}$ & $1.877^{\,0.048}$ & $1.021^{\,0.048}$ & $20.107^{\,0.018}$ & $5.973^{\,0.039}$ \\
$1{\rm c}I_{\rm \scriptscriptstyle C}$                           & $17.632^{\,0.030}$ & $1.957^{\,0.024}$ & $0.841^{\,0.028}$ & $19.217^{\,0.013}$ & $5.245^{\,0.022}$ \\
$2{\rm c}I_{\rm \scriptscriptstyle C}$                           & $17.977^{\,0.045}$ & $1.960^{\,0.034}$ & $1.083^{\,0.042}$ & $19.614^{\,0.011}$ & $6.443^{\,0.024}$ \\
$3{\rm c}I_{\rm \scriptscriptstyle C}$ {\rule[-1.0ex]{0pt}{0pt}} & $17.751^{\,0.057}$ & $2.036^{\,0.044}$ & $0.951^{\,0.049}$ & $19.356^{\,0.018}$ & $6.743^{\,0.049}$ \\

\hline \noalign {\smallskip}
\end{tabular}
\end{center}
\end{table}

\begin{table}
\begin{center}
\caption{\label{mags} Nucleus $(^{\rm \scriptscriptstyle N})$, bulge $(^{\rm \scriptscriptstyle B})$ and
disk $(^{\rm \scriptscriptstyle D})$ total magnitudes for III~Zw~2, Mrk~506 and Mrk~509. The errors of
the magnitudes are shown as a superscript index to the corresponding values. The values of
${\overline {\cal FW}}{\rm \scriptscriptstyle PSF}$ and $\sigma_{\rm fit}$ are
listed as well.}\smallskip
\begin{tabular}{lrrrrr}
\hline \noalign {\smallskip}
$\rm Code$ {\rule[-1.7ex]{0pt}{0pt}} & $\mu^{\,\rm \scriptscriptstyle N}_{\,\rm tot}$ & $\mu^{\,\rm \scriptscriptstyle B}_{\,\rm tot}$ & $\mu^{\,\rm \scriptscriptstyle D}_{\,\rm tot}$ & ${\overline {\cal FW}}{\rm \scriptscriptstyle PSF}$ & $\sigma_{\rm fit}$ \\
{\rule[-1.7ex]{0pt}{0pt}} & $[\,\rm mag\,]$ & $[\,\rm mag\,]$ & $[\,\rm mag\,]$ & $[\,^{\prime \prime}\,]$ & $[\,\rm mag\,/\,\Box^{\,\prime \prime}\,]$ \\
\hline \noalign {\smallskip}

$1{\rm a}B$                                                      & $16.935^{\,0.113}$ & $16.435^{\,0.193}$ & $17.831^{\,0.028}$ & $1.575$ & $0.025$ \\
$1{\rm a}V$                                                      & $16.073^{\,0.060}$ & $15.797^{\,0.143}$ & $16.214^{\,0.017}$ & $1.560$ & $0.017$ \\
$1{\rm a}R_{\rm \scriptscriptstyle C}$                           & $15.779^{\,0.186}$ & $15.222^{\,0.316}$ & $15.522^{\,0.051}$ & $1.311$ & $0.045$ \\
$1{\rm a}I_{\rm \scriptscriptstyle C}$ {\rule[-1.0ex]{0pt}{0pt}} & $14.893^{\,0.121}$ & $15.227^{\,0.341}$ & $14.512^{\,0.023}$ & $1.428$ & $0.047$ \\

\hline \noalign {\smallskip}

$2{\rm b}U$                                                      & $15.858^{\,0.120}$ & $15.446^{\,0.210}$ & $15.353^{\,0.053}$ & $1.816$ & $0.013$ \\ 
$1{\rm b}B$                                                      & $16.591^{\,0.010}$ & $               -$ & $15.046^{\,0.010}$ & $2.502$ & $0.028$ \\ 
$2{\rm b}B$                                                      & $16.725^{\,0.111}$ & $16.956^{\,0.244}$ & $15.158^{\,0.023}$ & $1.752$ & $0.035$ \\ 
$1{\rm b}V$                                                      & $16.314^{\,0.019}$ & $               -$ & $14.219^{\,0.007}$ & $2.350$ & $0.009$ \\ 
$2{\rm b}V$                                                      & $16.604^{\,0.082}$ & $16.150^{\,0.107}$ & $14.215^{\,0.016}$ & $1.516$ & $0.015$ \\ 
$1{\rm b}R_{\rm \scriptscriptstyle C}$                           & $16.230^{\,1.186}$ & $16.239^{\,2.963}$ & $13.674^{\,0.013}$ & $2.360$ & $0.014$ \\ 
$2{\rm b}R_{\rm \scriptscriptstyle C}$                           & $15.838^{\,0.023}$ & $15.838^{\,0.043}$ & $13.641^{\,0.010}$ & $1.753$ & $0.007$ \\ 
$1{\rm b}I_{\rm \scriptscriptstyle C}$                           & $15.690^{\,0.037}$ & $15.554^{\,0.050}$ & $12.914^{\,0.009}$ & $2.293$ & $0.007$ \\ 
$2{\rm b}I_{\rm \scriptscriptstyle C}$ {\rule[-1.0ex]{0pt}{0pt}} & $15.884^{\,0.037}$ & $15.077^{\,0.034}$ & $12.957^{\,0.014}$ & $1.459$ & $0.006$ \\ 

\hline \noalign {\smallskip}

$3{\rm c}U$                                                      & $12.664^{\,0.074}$ & $13.190^{\,0.201}$ & $15.459^{\,0.058}$ & $2.711$ & $0.037$ \\
$1{\rm c}B$                                                      & $15.288^{\,0.063}$ & $14.541^{\,0.086}$ & $16.134^{\,0.045}$ & $2.141$ & $0.020$ \\
$2{\rm c}B$                                                      & $15.191^{\,0.063}$ & $14.247^{\,0.091}$ & $15.813^{\,0.015}$ & $1.861$ & $0.016$ \\
$3{\rm c}B$                                                      & $14.140^{\,0.145}$ & $13.869^{\,0.212}$ & $15.696^{\,0.022}$ & $2.420$ & $0.024$ \\
$1{\rm c}V$                                                      & $14.595^{\,0.035}$ & $14.454^{\,0.066}$ & $15.004^{\,0.018}$ & $1.877$ & $0.031$ \\
$2{\rm c}V$                                                      & $14.398^{\,0.034}$ & $14.381^{\,0.082}$ & $14.799^{\,0.027}$ & $1.653$ & $0.022$ \\
$3{\rm c}V$                                                      & $14.169^{\,0.076}$ & $13.793^{\,0.144}$ & $14.901^{\,0.028}$ & $2.204$ & $0.014$ \\
$1{\rm c}R_{\rm \scriptscriptstyle C}$                           & $14.202^{\,0.020}$ & $13.874^{\,0.034}$ & $14.432^{\,0.011}$ & $1.777$ & $0.017$ \\
$2{\rm c}R_{\rm \scriptscriptstyle C}$                           & $13.920^{\,0.022}$ & $13.823^{\,0.049}$ & $14.379^{\,0.016}$ & $1.491$ & $0.023$ \\
$3{\rm c}R_{\rm \scriptscriptstyle C}$                           & $13.424^{\,0.023}$ & $13.816^{\,0.072}$ & $14.231^{\,0.018}$ & $2.252$ & $0.015$ \\
$1{\rm c}I_{\rm \scriptscriptstyle C}$                           & $13.929^{\,0.020}$ & $13.560^{\,0.030}$ & $13.623^{\,0.013}$ & $1.797$ & $0.019$ \\
$2{\rm c}I_{\rm \scriptscriptstyle C}$                           & $13.697^{\,0.019}$ & $13.785^{\,0.045}$ & $13.573^{\,0.011}$ & $1.977$ & $0.014$ \\
$3{\rm c}I_{\rm \scriptscriptstyle C}$ {\rule[-1.0ex]{0pt}{0pt}} & $13.289^{\,0.022}$ & $13.537^{\,0.057}$ & $13.216^{\,0.018}$ & $1.954$ & $0.017$ \\

\hline \noalign {\smallskip}
\end{tabular}
\end{center}
\end{table}

\begin{figure}[t]
\resizebox{\hsize}{!}{\includegraphics{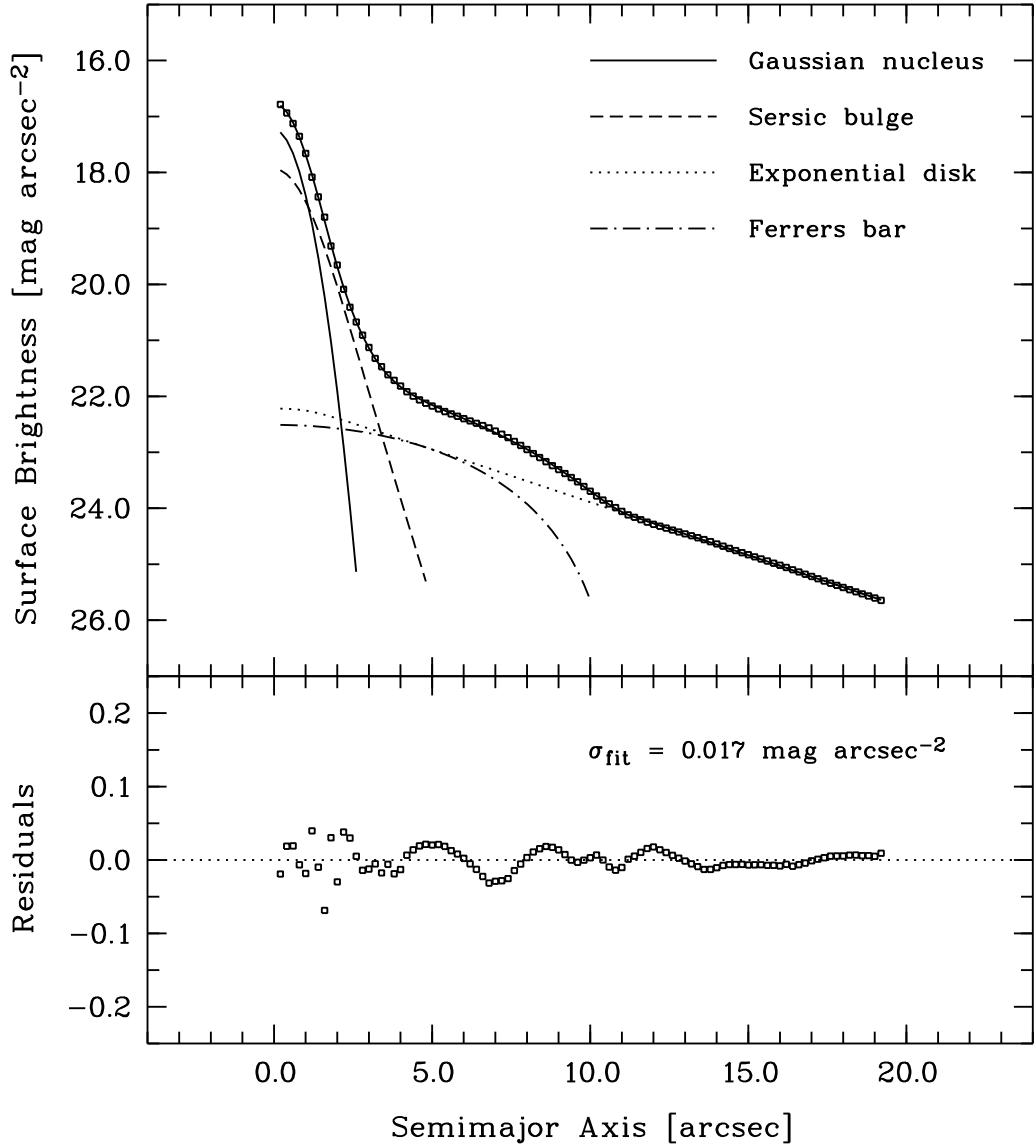}}
\caption{SBP decomposition for III~Zw~2. In the upper panel the observed SBP (open squares) and
the decomposed profiles of the structural components are shown; the resulting model SBP closely
follows the observed one. In the lower panel the residual profile is shown. \label{3zw2}}
\end{figure}

\begin{figure}[t]
\resizebox{\hsize}{!}{\includegraphics{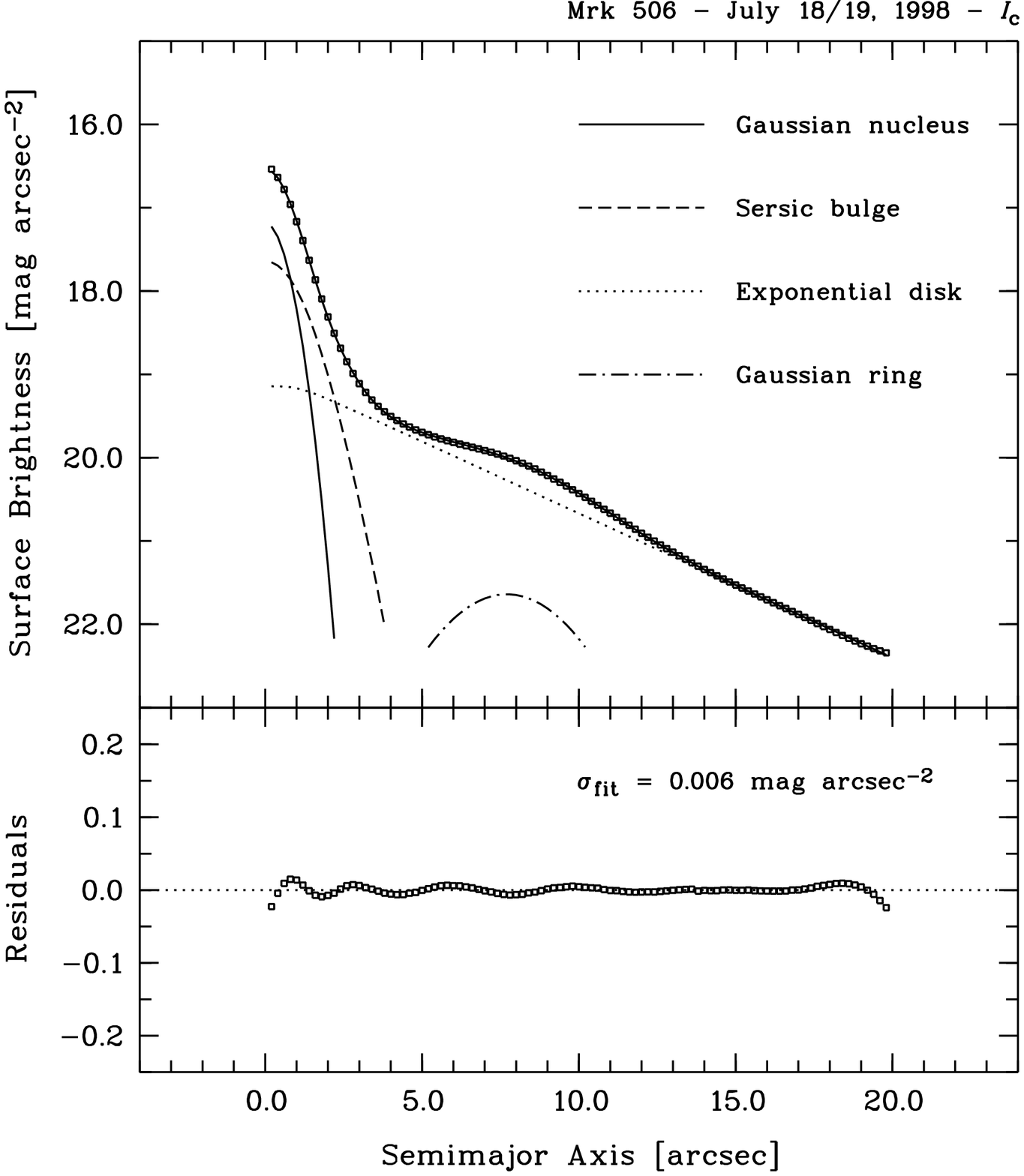}}
\caption{The same as in Fig~\ref{3zw2} but for Mrk~506. \label{m506}}
\end{figure}

\begin{figure}[t]
\resizebox{\hsize}{!}{\includegraphics{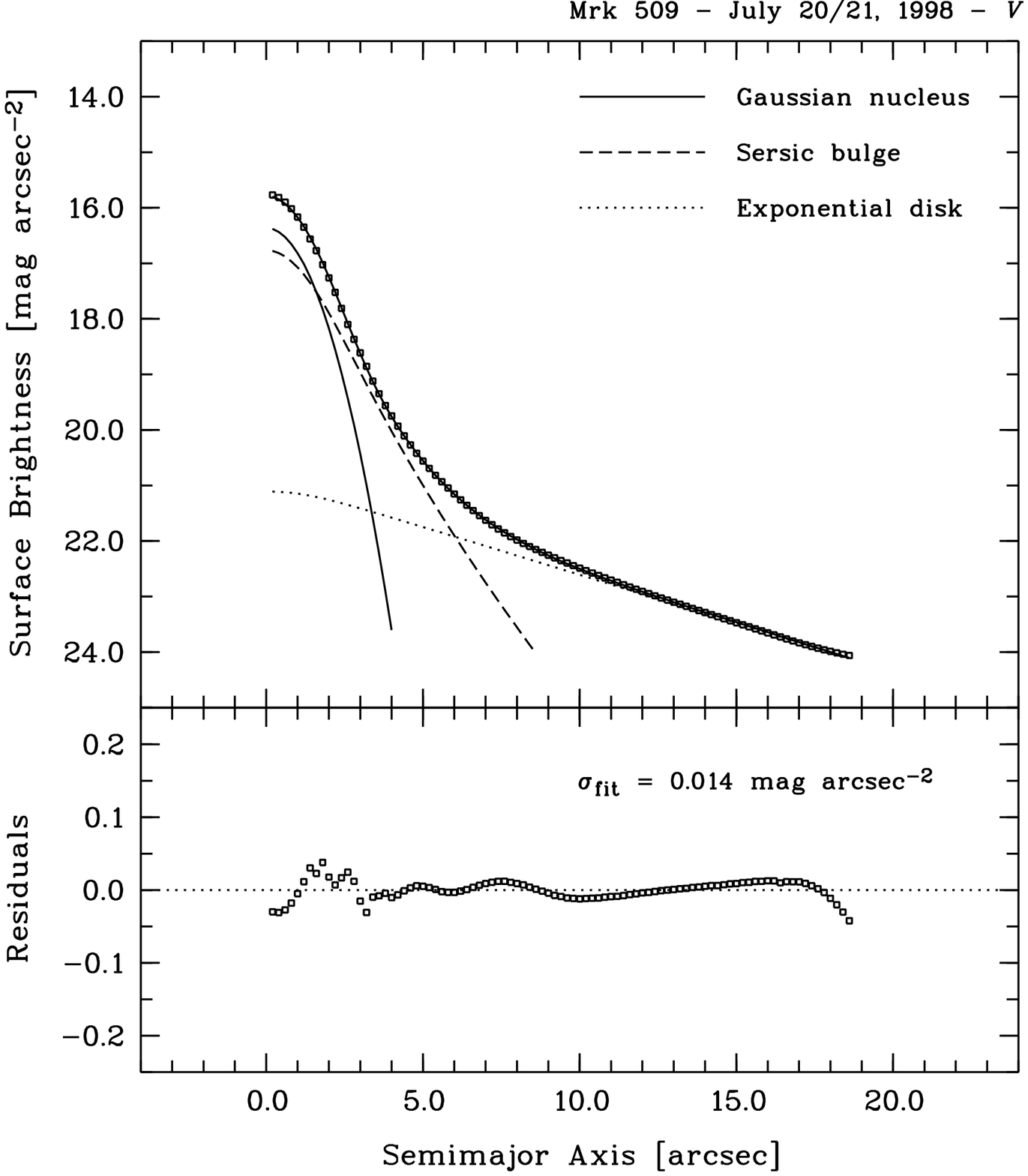}}
\caption{The same as in Fig~\ref{3zw2} but for Mrk~509. \label{m509}}
\end{figure}

The observed SBPs and the decomposed profiles of the structural components of III~Zw~2, Mrk~506
and Mrk~509 are shown in Fig.~\ref{3zw2}, Fig.~\ref{m506} and Fig.~\ref{m509}, respectively, along with
the residual profiles equal to $\mu^{\,\rm \scriptscriptstyle OBS}-\mu^{\,\rm \scriptscriptstyle MOD}(\textbf{\textit{p}}_{\rm min})$.
We show the decompositions with the smallest $\sigma_{\rm fit}$ among the different filters and
observing runs for each galaxy (see Table~\ref{mags}).

We list in Table~\ref{bar} the length of the bar in III~Zw~2 obtained from the
$BVR_{\rm \scriptscriptstyle C}I_{\rm \scriptscriptstyle C}$ ellipticity profiles using (1)
a maximum ellipticity criterion -- the bar length, $l^{\,\rm (max)}_{\rm bar}$, corresponds
to the point of maximal ellipticity, and (2) a minimum ellipticity criterion -- the bar
length, $l^{\,\rm (min)}_{\rm bar}$, corresponds to the point of minimal ellipticity next to
the ellipticity maximum, so one has $l^{\,\rm (min)}_{\rm bar}>l^{\,\rm (max)}_{\rm bar}$ by definition.
One could see that $l^{\,\rm (min)}_{\rm bar}$ agree well with the bar length, $r_{\,\rm end}$,
obtained from the SBPs decomposition, while $l^{\,\rm (max)}_{\rm bar}$ underestimates $r_{\,\rm end}$
(cf. \cite{wozniak95}).

$BVI_{\rm \scriptscriptstyle C}$ profiles decomposition of Mrk~509 was recently presented by \cite{boris02}
where a sum of a Gaussian, a de~Vaucouleurs and a truncated exponent was used as a model SBP. We have not
found evidence of Freeman type II profile in Mrk~509 to justify the usage of a truncated exponent. We have
found a nearly exponential bulge, $n \approx 1$, in all three galaxies decomposed (see also \cite{balcells03,aguerri05}).

\begin{table}
\begin{center}
\caption{\label{bar} Bar parameters for III~Zw~2. The bar lengths obtained from the ellipticity profiles employing
the maximal and the minimal ellipticity criteria are listed in the last two columns for comparison
with the bar length obtained from the SBPs decomposition.}\smallskip
\begin{tabular}{lrrrrr}
\hline \noalign {\smallskip}
$\rm Code$ {\rule[-1.7ex]{0pt}{0pt}} & $\mu_{\,\rm cen}$ & $r_{\,\rm end}$ & $m$ & $l^{\,\rm (max)}_{\rm bar}$ & $l^{\,\rm (min)}_{\rm bar}$ \\
{\rule[-1.7ex]{0pt}{0pt}} & $[\,\rm mag\,/\,\Box^{\,\prime \prime}\,]$ & $[\,^{\prime \prime}\,]$ & & $[\,^{\prime \prime}\,]$ & $[\,^{\prime \prime}\,]$ \\
\hline \noalign {\smallskip}

$1{\rm a}B$                                                      & $23.941^{\,0.043}$ & $11.430^{\,0.142}$ & $1.874^{\,0.116}$  & $7.978$ & $11.350$ \\
$1{\rm a}V$                                                      & $22.510^{\,0.034}$ & $11.258^{\,0.110}$ & $1.852^{\,0.091}$  & $6.991$ & $11.214$ \\
$1{\rm a}R_{\rm \scriptscriptstyle C}$                           & $22.189^{\,0.111}$ & $11.906^{\,0.701}$ & $2.506^{\,0.536}$  & $6.726$ & $12.145$ \\
$1{\rm a}I_{\rm \scriptscriptstyle C}$ {\rule[-1.0ex]{0pt}{0pt}} & $22.409^{\,0.140}$ & $ 9.468^{\,0.213}$ & $0.876^{\,0.190}$  & $6.824$ & $11.252$ \\

\hline \noalign {\smallskip}
\end{tabular}
\end{center}
\end{table}

\begin{table}
\begin{center}
\caption{\label{ring} Ring parameters for Mrk~506.}\smallskip
\begin{tabular}{lrrr}
\hline \noalign {\smallskip}
$\rm Code$ {\rule[-1.7ex]{0pt}{0pt}} & $\mu_{\,\rm cen}$ & $r_{\,\rm cen}$ & $\cal FW$  \\
{\rule[-1.7ex]{0pt}{0pt}} & $[\,\rm mag\,/\,\Box^{\,\prime \prime}\,]$ & $[\,^{\prime \prime}\,]$ & $[\,^{\prime \prime}\,]$ \\
\hline \noalign {\smallskip}

$2{\rm b}U$                                                      & $24.119^{\,0.079}$ & $8.679^{\,0.118}$ & $4.874^{\,0.240}$ \\
$1{\rm b}B$                                                      & $24.140^{\,0.053}$ & $8.793^{\,0.086}$ & $3.430^{\,0.199}$ \\
$2{\rm b}B$                                                      & $23.750^{\,0.049}$ & $8.721^{\,0.098}$ & $4.035^{\,0.224}$ \\
$1{\rm b}V$                                                      & $22.754^{\,0.013}$ & $8.211^{\,0.027}$ & $4.926^{\,0.059}$ \\
$2{\rm b}V$                                                      & $22.999^{\,0.037}$ & $8.118^{\,0.073}$ & $4.656^{\,0.163}$ \\
$1{\rm b}R_{\rm \scriptscriptstyle C}$                           & $22.451^{\,0.033}$ & $8.755^{\,0.086}$ & $4.356^{\,0.143}$ \\
$2{\rm b}R_{\rm \scriptscriptstyle C}$                           & $22.167^{\,0.019}$ & $7.685^{\,0.041}$ & $5.392^{\,0.083}$ \\
$1{\rm b}I_{\rm \scriptscriptstyle C}$                           & $21.469^{\,0.015}$ & $8.273^{\,0.043}$ & $4.959^{\,0.073}$ \\
$2{\rm b}I_{\rm \scriptscriptstyle C}$ {\rule[-1.0ex]{0pt}{0pt}} & $21.640^{\,0.029}$ & $7.705^{\,0.057}$ & $5.443^{\,0.110}$ \\

\hline \noalign {\smallskip}
\end{tabular}
\end{center}
\end{table}

\section*{Acknowledgments}

This research has made use of the NASA/IPAC Extragalactic Database (NED) which is
operated by the Jet Propulsion Laboratory, California Institute of Technology, under
contract with the National Aeronautics and Space Administration.

\end{document}